# A spin-down clock for cool stars from observations of a 2.5-billion-year-old cluster

Søren Meibom[1], Sydney A. Barnes[2], Imants Platais[3], Ronald L. Gilliland[4], David W. Latham[1], Robert D. Mathieu[5]

[1] Harvard-Smithsonian Center for Astrophysics, Cambridge, MA 02138 USA.

[2] Leibniz-Institute for Astrophysics, An der Sternwarte 16, 14482 Potsdam, Germany / Space Science Institute, USA

[3] Johns Hopkins University, 3400 North Charles Street, Baltimore, MD 21218, USA

[4] Center for Exoplanets and Habitable Worlds, The Pennsylvania State University, University Park, PA 16802, USA

[5] University of Wisconsin - Madison, Madison, WI 53706, USA

**The ages of the most common stars—low-mass (cool) stars like the Sun, and smaller—are difficult to derive[1,2] because traditional dating methods use stellar properties that either change little as the stars age[3,4] or are hard to measure[5–8]. The rotation rates of all cool stars decrease substantially with time as the stars steadily lose their angular momenta. If properly calibrated, rotation therefore can act as a reliable determinant of their ages based on the method of gyrochronology[2,9–11]. To calibrate gyrochronology, the relationship between rotation period and age must be determined for cool stars of different masses, which is best accomplished with rotation period measurements for stars in clusters with well-known ages. Hitherto, such measurements have been possible only in clusters with ages of less than about one billion years[12–16], and gyrochronology ages for older stars have been inferred from model predictions[2,7,11,17]. Here we report rotation period measurements for 30 cool stars in the 2.5-billion-year-old cluster NGC 6819. The periods reveal a well-defined relationship between rotation period and stellar mass at the cluster age, suggesting that ages with a precision of order 10 per cent can be derived for large numbers of cool Galactic field stars.**

Prior observations in star clusters with ages ≲300 million years (Myr) have shown that cool stars begin their main-sequence phase with a dispersion in their rotation periods, $P$, spanning two orders of magnitude[12–14,18] (0.1–10 d). However, this dispersion diminishes rapidly with cluster age, $t$, as they lose angular momentum through magnetically channelled winds[19], causing their periods to increase and converge to a well-defined relationship with stellar mass, $M$, by the age (600 Myr) of the Hyades cluster[15,20]. These observations suggest that cool main-sequence stars older than the Hyades probably occupy a single surface, $P = P(t, M)$, in $P$–$t$–$M$ space, which can be defined by measurements of their periods and photometric colours (a proxy for stellar mass) in a series of age-ranked clusters (Fig. 1). Measuring stellar rotation periods in clusters older than the Hyades will confirm or deny the existence of such a surface, and, if it exists, define its shape and thickness.

Models of cool-star rotational evolution also describe a convergence of rotation with age[2,11,17,21]. However, they differ in their predictions of the location and shape of the $P$–$t$–$M$ surface and, beyond the common assumption that the rotation period of the Sun is typical for stars of its mass and age, were until 2011[16] unconstrained at ages greater than 600 Myr.

Observations to define the $P$–$t$–$M$ surface at older ages are therefore required to extend our knowledge of the $P$–$t$–$M$ relations and thereby allow the ages of individual cool stars in the field to be derived from their measured periods and colours by the method of gyrochronology[11].

The rotation period of a cool star can be determined from small (≤1%) periodic modulations in its brightness as rotation carries star-spots across the stellar disc. Older stars have fewer and smaller spots, making their periods harder to detect. Accordingly, observations from ground-based telescopes have been unable to detect rotation periods in clusters older than the Hyades. Although periods are increasingly being measured for isolated field stars[22], their ages, unlike those of cluster stars, are not known to a precision adequate to calibrate gyrochronology.

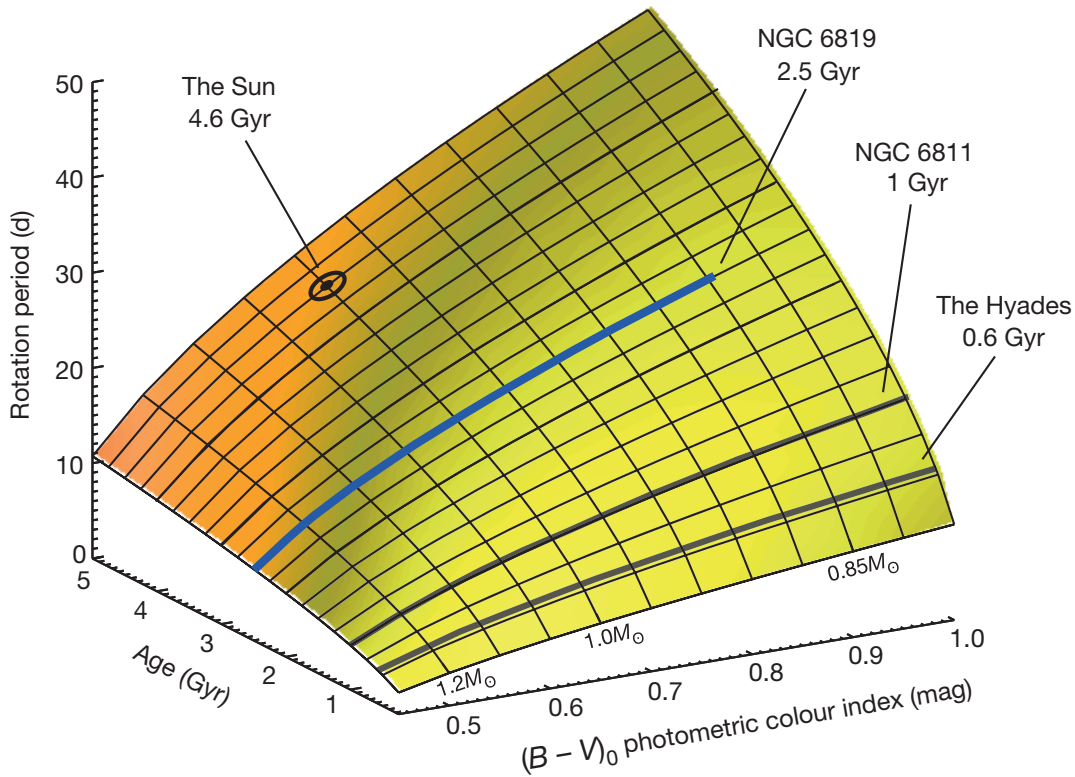

**Figure 1 | The schematic $P$–$t$–$M$ surface for cool stars.** The hypothetical relationship between rotation period, age and colour extrapolated (yellow) to greater ages from the colour–period relations in young clusters using a particular $P$–$t$ relationship[5], and assuming that the Sun (marked by the black solar symbol; ☉) resides on it. The blue line indicates the locus of stars in NGC 6819 for which we have determined rotation periods. The dark grey lines at ages of 0.6 and 1 Gyr represent prior observations in the Hyades[15] and NGC 6811[16] clusters, respectively. Stellar masses in solar units are marked on the surface at the corresponding colours. (Figure adapted from ref. 16.)

With an age of 2.5 billion years (Gyr) (ref. 23), NGC 6819 bridges the large gap in age between the Sun and existing cluster observations (Fig. 1). The cluster was within the field of view of NASA's Kepler satellite, permitting a time-series photometric survey of its members by The Kepler Cluster Study[16]. Cool cluster members were selected for observation using prior ground-based photometry for the cluster[24], a 90 yr-baseline proper-motion study[25] and multi-epoch radial-velocity measurements over 15 yr (ref. 26 and an ongoing survey by S.M.) (Me-thods). The precision, cadence and duration of the Kepler photometry enable us to measure rotation periods for cool stars much older than the Hyades. Previous results from The Kepler Cluster Study in the 1 Gyr-old cluster NGC 6811 confirmed the existence of a unique relationship between $P$ and $M$ for cool stars at that age[16], and measured a median rotation period of 10.8 d for solar-mass stars. Since then, a study to measure periods from Kepler data in NGC 6819 was carried out by another group[27]. That work was limited to cluster stars of greater than solar mass, and was thus unable to define a $P$–$M$ relationship for cool stars.

We have measured rotation periods for 30 cool stars in NGC 6819 (Extended Data Figs 1-5). All 30 stars are both photometric (from their location in the cluster's colour–magnitude diagram) and kinematic (from measurements of their proper motions and radial velocities) members of the cluster (Methods). Their de-reddened colour indices, $(B-V)_0$, range from 0.41 to 0.89 mag, corresponding to a stellar mass range from ~1.4 to 0.85 solar masses. Their periods range from 4.4 to 23.3 d and are displayed in the resulting colour–period diagram (CPD) for NGC 6819 (Fig. 2). Extended Data Table 1 lists all relevant properties for the 30 stars. The stars form a single and narrow sequence in the CPD. This sequence defines a clear dependence of increasing stellar rotation period (decreasing rotation rate) on increasing stellar colour (decreasing mass) and represents the cross-section of the hypothesized $P$–$t$–$M$ surface at t = 2.5 Gyr (Fig. 1, blue line).

The solar-mass stars (defined here as those with 0.62 mag ≤ $(B-V)_0$ ≤ 0.68 mag) all have periods between 17.36 and 18.70 d (mean, 18.2 d; s.d., 0.4 d), implying that the Sun's rotation period was probably in that range when it was the age of NGC 6819. With the 10.8 d median period for solar-mass stars in NGC 6811[16], and the mean solar photometric rotation period of 26.1 d, this implies a Skumanich-type[5] spin-down ($P$ varies as $t^{1/2}$) for solar-mass stars over the 3.6 Gyr interval measured.

The relatively small number of periods detected for stars with $(B-V)_0$ from 0.47 to 0.57 mag does not reflect a lack of cluster members (Extended Data Fig. 6). A similar pattern was seen for NGC 6811[16], and is unsurprising to photometrists, who know that such stars show little variability. The colour index $(B-V)_0 = 0.47$ mag separates stars with radiative envelopes from those with convective envelopes and is associated with the onset of effective magnetic wind braking (the 'break in the Kraft curve'[28]). The rotation periods of the more massive stars in NGC 6819 ($(B-V)_0 < 0.47$ mag) are scattered around a median of 4.8 d, demonstrating a steeper spin-down ($P$ varies as $t$) from a median period of 1.3 d in the 1 Gyr-old cluster NGC 6811[16].

The 30 rotation periods were determined by Lomb-Scargle periodogram analysis[29] of long-cadence (30 min exposures) Kepler light curves spanning ~3.75 yr (Methods). The rotation period for a given star was determined from subsections of the full light curve, chosen to avoid and minimize the effect on the measured period of multiple spots or spot groups, or that of trends not removed by the data processing, or both. For all periods reported we have manually examined the periodogram and the phased and unphased light curves, determined the periods independently using the CLEAN algorithm[30], and assessed the level of contamination from neighbouring stars (Methods and Extended Data Fig. 7).

We have also measured the projected rotation velocities ($v\sin(i)$, where $v$ is the stellar rotation velocity and $i$ is the inclination angle between the stellar spin axis and the observer's line of sight) spectroscopically for 25 of the 30 stars. These $v\sin(i)$ values are fully consistent with the photometric rotation periods (Extended Data Fig. 8). The resolution of the spectra for the five remaining stars is too low to provide meaningful constraints on their rotation velocities.

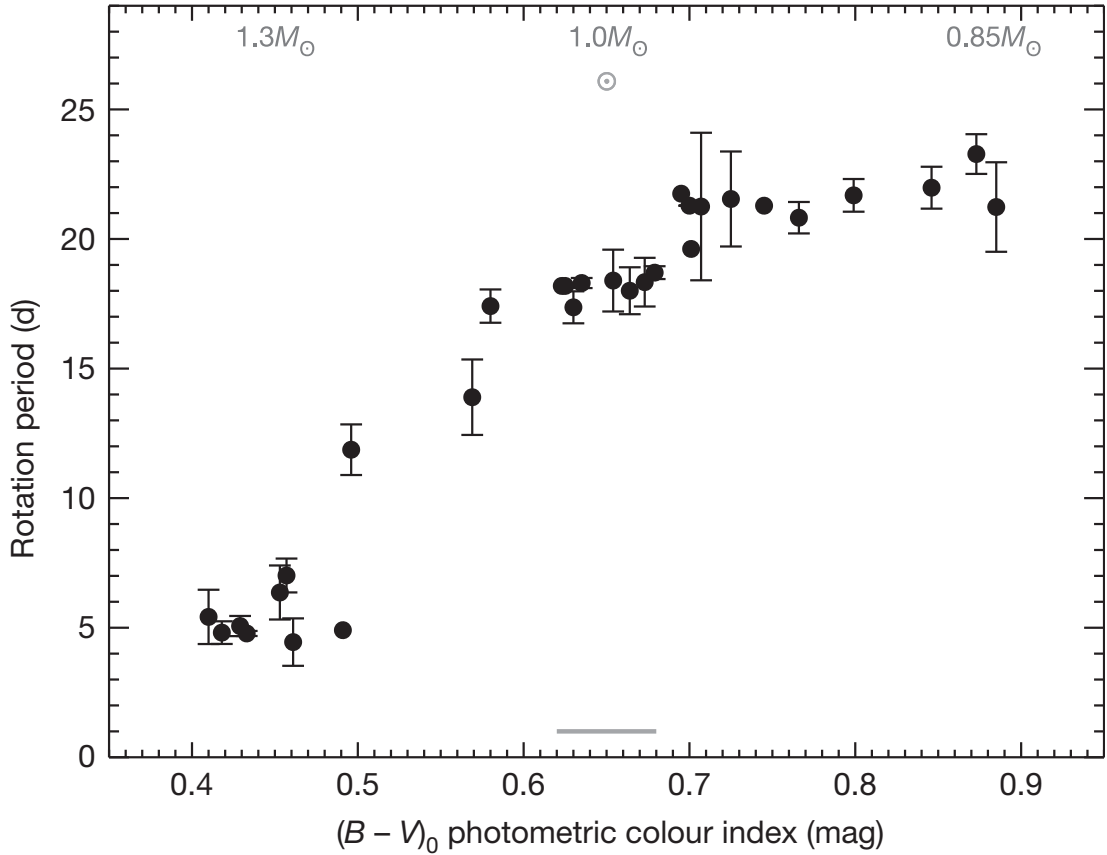

**Figure 2 | The colour–period diagram for NGC 6819.** The distribution of rotation periods as a function of de-reddened colour index $(B-V)_0$ for 30 cool photometric, proper-motion, and radial-velocity members of the 2.5 Gyr open star cluster NGC 6819. The measurements define a tight dependence of rotation period on colour (mass). The symbols and error bars respectively indicate the means and standard deviations of multiple measurements for the same star when available. The location of the Sun (4.56 Gyr) in the diagram is marked with a grey solar symbol. Stellar masses in solar units are given along the top horizontal axis at the corresponding colours. Solar-mass stars with $(B-V)_0$ between 0.62 and 0.68 mag (interval marked by grey line near the bottom horizontal axis) have a mean period of 18.2 d with a standard deviation of 0.4 d.

The measured rotation periods in NGC 6819 establish that the $P$–$t$–$M$ surface is well defined at ages beyond 1 Gyr. Together with prior observations in younger clusters, they specify the location, shape and thickness of the surface to an age of 2.5 Gyr.

The CPD is the projection of this surface onto the colour-period plane. Therefore, in this diagram, the measured rotational sequence for NGC 6819 separates stars younger (below the sequence) and older (above the sequence) than 2.5 Gyr, independent of any theoretical model of stellar rotational evolution. Because angular momentum loss in cool single stars is driven by internal processes independent of environment (pathological tidally interacting systems are an exception to this, but tidal effects on rotation are not a concern for ~97% of cool field stars; see Methods), it is the same for both cluster and field stars. Therefore, this classification, and, more generally, the relationship between rotation and age, must also be valid for cool field stars.

The relative scatter (dP/P) about the NGC 6819 rotational sequence is ~10% for our entire sample, ~5% for stars with $(B-V)_0 > 0.55$ mag and ~2% (0.4 d) for the domain surrounding solar-mass stars that we are able to define particularly well. This scatter includes contributions from period measurement uncertainties, and the residual effects of both stellar differential rotation and the spread in initial periods on the 'zero-age main sequence'. Its relatively small size demonstrates that these effects do not prevent the determination of age from rotation, and that the $P$–$t$–$M$ surface is intrinsically thin at this age, implying that ages determined from spin-down ('gyro ages') will be precise.

The derivation of such ages requires a model. A number of such models exist[2,7,11,17], individually differing with respect to the functional forms of the underlying variables, and even with respect to what those specific variables are. The NGC 6819 rotation period data permit a comparison between the predictions of these rotational evolution models for its age and the actual measurements. Figure 3 shows this comparison.

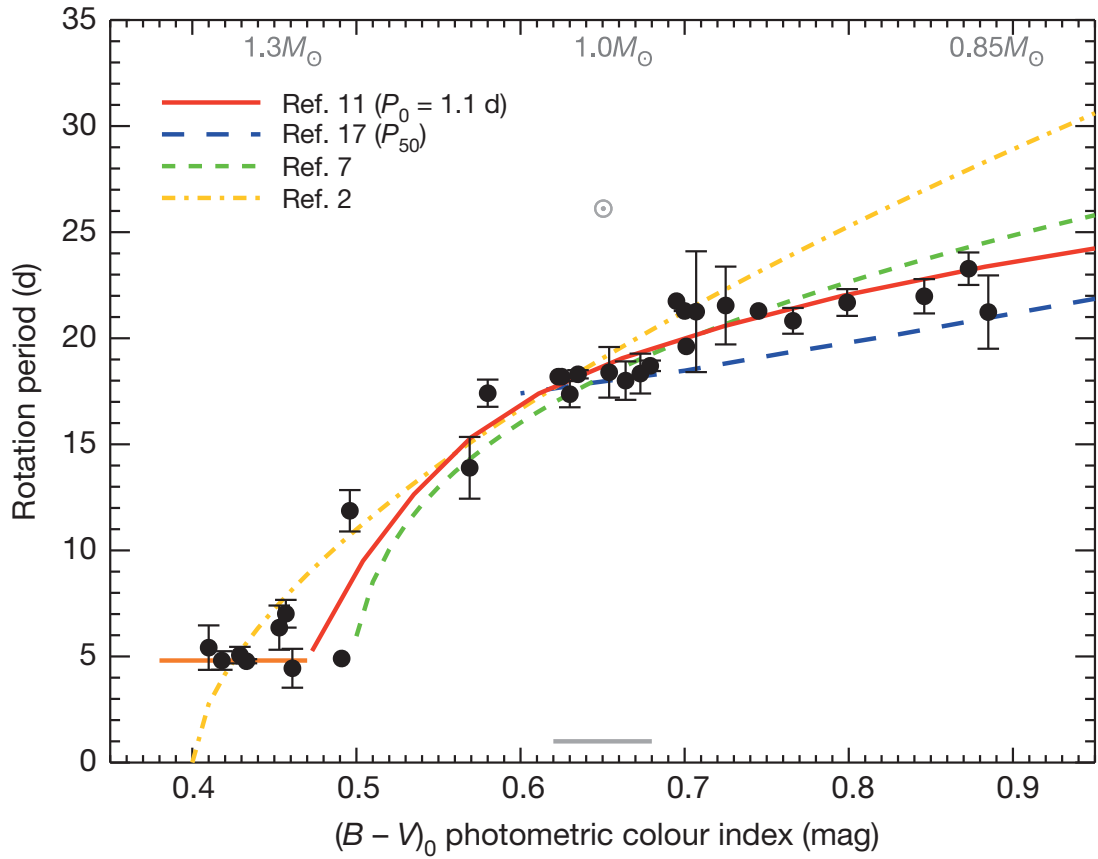

**Figure 3 | Comparison between gyrochronology models and the NGC 6819 CPD.** The predictions from four different models[2,7,11,17] of cool star rotation periods at 2.5 Gyr are plotted against the measured periods in NGC 6819. All plotted models predict an observed increase in rotation period with increasing (B 2 V)0 colour (decreasing stellar mass). The colour–period relation from ref. 11 fits the observations for stars with $(B-V)_0 > 0.55$ mag. The symbols and error bars respectively indicate the means and standard deviations of multiple measurements for the same star when available. The location of the Sun (4.56 Gyr) is marked with a grey solar symbol. Stellar masses in solar units are given along the top horizontal axis at the corresponding colours. The colour range for solar-mass stars is marked with a solid grey line near the bottom horizontal axis. The orange horizontal line for $(B-V)_0 < 0.47$ mag marks the median period of 4.8 d for stars in this colour range (~1.2–1.4 solar masses). $P_0$ in the model from ref. 11 refers to the initial (zero-age main sequence) rotation period. $P_{50}$ in the model from ref. 17 refers to the 50th percentile rotation period for a given stellar mass.

The (unaltered) model of gyrochronology from ref. 11 provides a good fit to the data over nearly the full colour range of the observations. To test the precision of gyrochronology we may thus treat the coeval NGC 6819 stars as individual field stars, and ask what age the model of ref. 11 would provide for each of the 21 best-measured stars, that is, those with $(B-V)_0$ colour index between 0.55 and 0.9 mag (masses between ~1.1 and ~0.85 that of the Sun). Every one of these stars returns a gyro age between 2 and 3 Gyr, with a roughly Gaussian distribution centred at 2.49 Gyr (Extended Data Fig. 9). The standard deviation of the 21 ages is 0.25 Gyr (10% of the mean gyro age), implying that ages of this precision can be derived for similarly well-measured field stars, despite the effects of measurement errors, differential rotation and a spread in initial rotation periods.

The mean age of 2.49 Gyr also represents the gyro age for NGC 6819. The standard error in this cluster age is 0.056 Gyr (that is, a 2% uncertainty, ignoring possible systematic errors in gyrochronology). The cluster gyro age thus agrees to within the uncertainty with the classical stellar evolution age of the cluster[23], implying that gyrochronology is well-calibrated at 2.5 Gyr.

We conclude that gyrochronology can provide accurate and precise ages for large numbers of cool stars with measured rotation periods. Such ages will enable us to study how astrophysical phenomena involving cool stars evolve over time, and will therefore be important to a wide range of research from the Galactic scale down to the scale of individual stars and their companions.

**Acknowledgements** S.M. acknowledges support through NASA grant NNX09AH18A (The Kepler Cluster Study), NSF grant 1312882 (The Kepler Cluster Study: Planets and Gyrochronology), and the Smithsonian Institution's Competitive Grants Program for Science in 2012 and 2013. SAB acknowledges generous support from the German Science Foundation (DFG) during a crucial phase of this work via a Mercator Guest Professorship at the University of Potsdam and the Leibniz Institute for Astrophysics Potsdam, Germany. This paper includes data collected by the Kepler mission. Kepler was competitively selected as the tenth Discovery mission. Funding for the Kepler mission is provided by the NASA Science Mission directorate. Some/all of the data presented in this paper were obtained from the Mikulski Archive for Space Telescopes (MAST). STScI is operated by the Association of Universities for Research in Astronomy, Inc., under NASA contract NAS5-26555. Support for MAST for non-HST data is provided by the NASA Office of Space Science via grant NNX13AC07G and by other grants and contracts. Spectroscopic observations in NGC 6819 with the Hectochelle spectrograph were obtained at the MMT Observatory, a joint facility of the Smithsonian Institution and the University of Arizona.

**Author Contributions** SM is the Principal Investigator for The Kepler Cluster Study and led the planning and execution of the study, the determination and validation of rotation periods from the *Kepler* data, the membership, binarity, and *vsini* survey in NGC6819 with the Hectochelle spectrograph on the MMT telescope, and the writing of this paper. SAB is Co-Investigator on The Kepler Cluster Study and participated in planning of the study, evaluated the light curves for periodicity, performed the gyrochronology analysis and ages for the stars, and collaborated closely with SM in writing the paper. IP is Co-Investigator on The Kepler Cluster Study and contributed to the selection of Kepler targets in NGC 6819 with proper motion membership information and to the analysis of crowding and contamination in the vicinity of target stars from deep and high-resolution images of this star cluster. RLG is Co-Investigator on The Kepler Cluster Study and contributed to the selection of Kepler targets in NGC 6819 with analysis of crowding and contamination in deep high spatial resolution images. He carried out analysis of alternative methods for extracting light curves from raw Kepler data. DWL participated in the NGC 6819 radial-velocity membership and binarity surveys and lead the preparation of the Kepler Input Catalog (KIC). RDM is the Principal Investigator of the WIYN Open Cluster Study (WOCS) which contributed radial-velocity measurements for the membership and binary star surveys in NGC 6819.

# METHODS

## 1. Cluster membership for the 30 stars with measured rotation periods

### 1.1. Radial-velocity membership.

The common space motion of the stars in a cluster is an effective way to distinguish them from foreground or background stars in the Galactic disk. Using the Hydra and Hectochelle multi-object spectrographs on the WIYN 3.5 m and MMT 6.5 m telescopes we have measured radial (line-of-sight) velocities over 15 yr for more than 4,300 stars within a circular 1 degree diameter field centred on NGC 6819 (ref. 26 and an ongoing survey by S.M.). The Hydra spectra cover a 25 nm window centred on 513 nm and have a resolution of ~20,000. The Hectochelle spectra have resolutions of ~40,000 over a 15 nm window centred at 522 nm. For the late-type stars in NGC 6819 these spectral ranges are rich with narrow absorption lines and are thus well suited for radial-velocity measurements. Our radial-velocity measurement precision for stars of spectral types F, G and K (masses from ~1.3 to 0.7 solar masses) is ~0.4 kms$^{-1}$ for stars brighter than 18.5 mag in the V band.

Against the broad radial-velocity distribution of Galactic field stars in the direction of the cluster, the members of NGC 6819 populate a distinct peak with a mean radial velocity of +2.6 ± 0.8 kms$^{-1}$. The uncertainty represents the velocity dispersion among the stars caused by internal cluster dynamics, binary orbital motions and observational errors. For a given star, the probability of cluster membership ($P_{RV}$) is calculated from simultaneous fits of separate Gaussian functions to the cluster (FC) and field (FF) radial-velocity distributions. The probability is defined as the ratio of the cluster-fitted value over the sum of the cluster- and field-fitted values at the star's radial velocity[26] (RV):

$$P_{RV} = FC(RV) / [FC(RV) + FF(RV)]$$

The 30 stars in NGC 6819 with measured rotation periods are all radial-velocity members of the cluster ($P_{RV}$ > 50%) and their membership probabilities are given in Extended Data Table 1. The radial-velocity measurements also suggest that none of the 30 stars are in short-period binary stars, and, thus, that their angular momentum evolution is not affected by tidal interactions[31,32].

### 1.2. Proper-motion membership.

For NGC 6819, a number of archival photographic plates are available, taken between 1919 and 1973 with long-focus telescopes. A total of 23 photographic plates were digitized using the Space Telescope Science Institute's GAMMA II multi-channel microdensitometer[33]. Combining the measurements of these plates with matching second-epoch high-spatial-resolution CCD images, obtained in 2009 with the MegaCam camera on the 3.6 m Canada-France Hawaii Telescope (CFHT), allows us to derive accurate proper motions for stars in the field of NGC 6819. Proper motions were calculated for 15,750 stars down to 22 mag in V over a 40 arcmin by 40 arcmin field centred on NGC 6819. The accuracy of the proper motions for well-measured stars is ~0.2 mas yr$^{-1}$. For the inner parts (within a 15 arcmin radius from the centre) of the cluster this accuracy holds for stars brighter than 18 mag in V. Considering that the intrinsic dispersion of field star proper motions in the direction of NGC 6819 is about 3 mas yr$^{-1}$, this high accuracy enables a clean separation of cluster stars from field stars. Cluster membership probabilities ($P_\mu$) were calculated using

$$P_\mu = \Phi_{\text{Cluster}}/(\Phi_{\text{Cluster}} + \Phi_{\text{Field}})$$

where $\Phi_{\text{Cluster}}$ and $\Phi_{\text{Field}}$ are the two-dimensional Gaussian frequency distribu- tions of the cluster and field stars, respectively[25].

The majority (21) of the 30 stars with measured rotation periods have $P_\mu > 90\%$. For the remaining stars $P_\mu$ is lower because they are either located in the periphery of the cluster or their proper-motion errors are higher than expected at their apparent magnitudes[25]. The latter usually is due to some degree of stellar image overlap.

### 1.3 Photometric membership.

The colour–magnitude diagram (CMD; Extended Data Fig. 6) provides a third set of criteria for cluster membership. In the CMD, cluster members trace a well-defined relationship between stellar mass (B-V colour index) and luminosity (brightness, V). Extended Data Fig. 6 shows V and (B-V)$_0$ for proper-motion members of NGC 6819. The cluster members form a clearly visible diagonal band in the CMD and a 'hockey-stick'-like turn-off near (B-V)$_0 \sim 0.45$ mag and V $\sim 16$ mag. The locations of the 30 stars with measured rotation periods are marked with larger red circles. They are located on the cluster band in the CMD, making them photometric members of NGC 6819. The two members near (B-V)$_0 = 0.49$ mag are photometric binaries. The combined light from the two stars in the binary system places them above the cluster sequence. For these two stars, three and four radial-velocity measurements over 1,732 d and 1,074 d, respectively, show radial-velocity variations near our measurement precision, suggesting that these stars are members of relatively wide binary systems.

### 2. Data and data analysis.

Stars identified as members of NGC 6819 were added to the list of targets for the Kepler mission as part of The Kepler Cluster Study[16]. The 30 stars for which we measure rotation periods were observed by Kepler for 2.5 yr, on average, over ~3.75 yr. NGC 6819 was located in the part of the Kepler field of view covered by the CCD module that failed in January 2010. Since then, targets in NGC 6819 could be observed for only three of the four quarters each year.

Stellar rotation periods were derived from Kepler data summed into long-cadence (~30 min) bins. The data were processed by version 8.0 of the Kepler mission's data analysis pipeline and corrected by the Kepler Presearch Data Conditioning (PDC) module of the pipeline with an additional Bayesian maximum a posteriori (MAP) approach for the removal of systematics while preserving astrophysical signals such as rotational modulation[34,35]. Before performing our period search, all quarters of the corrected data were normalized by the median signal and joined together to form a single light curve for each star. We used Lomb–Scargle periodogram analysis[29] to detect periodic variability, searching 20,000 frequencies corresponding to periods between 0.05 and 100 d. The rotation period for a given star was determined from between one and seven separate time intervals distributed over the full light curve. The median peak-to-peak amplitude of variability for all period detections is 4 mmag, with a range of 1 to 125 mmag. When more than one rotation period measurement was possible for a given star, the mean period was calculated and used. For all stars, the periodogram and the raw and phased light curves were examined by eye. The periods were derived independently by both S.M. and S.A.B. using different analysis tools and algorithms. Extended Data Figs 1–5 shows examples, for all 30 stars, of PDC-MAP corrected light curve intervals used to measure their rotation periods. The figure also shows the light curves phase-folded on the periods, and the corresponding periodogram for each star. Extended Data Table 1 lists, for each of the 30 stars, basic astrometric and photometric data, the radial-velocity and proper-motion cluster membership probabilities, the number of period measurements, and their period mean and standard deviation.

For 25 of the 30 stars, spectra acquired with Hectochelle enabled determination of their projected rotation velocities ($v\sin(i)$) via cross-correlation with a library of synthetic spectra. Extended Data Fig. 8 shows the mean rotation periods versus the mean $v\sin(i)$ for stars in NGC 6819. It also displays three curves tracing the expected relation between rotation period and rotation velocity for a 90 degree inclination angle ($i$) of the stellar rotational axis and stellar radii of 0.85, 1.0, 1.4 solar radii. The error bars represent the standard deviation of multiple rotation period and $v\sin(i)$ measurements. The figure demonstrates that the photometrically measured rotation periods are consistent with the spectroscopically derived rotation velocities.

## 3. The distribution of gyrochronology ages for the 30 NGC 6819 members.

We have calculated the gyro ages for 21 stars with $(B-V)_0$ between 0.55 and 0.9 mag (masses between ~1.1 and 0.85 solar masses) using the model of ref. 11. Although this model is valid for stars with $(B-V)_0 > 0.47$ mag, the NGC 6819 rotation sequence is poorly defined (by only two stars) for 0.47 mag < $(B-V)_0$ < 0.55 mag.

For each of the 21 stars we have converted its $(B-V)_0$ colour into a value of the (global) convective turnover timescale, $t$, using a numerical one-to-one transformation table[36], and linear interpolation as needed. The resulting $t$ values are associated with the measured rotation period values, $P$, and inserted into equation (32) in ref. 11, that is

$$t = (\tau/k_C) \ln(P/P_0) + (k_I/2\tau)(P^2 - P_0^2)$$

using an initial period of $P_0 = 1.1$ d, as suggested in ref. 11. The dimensionless constants $k_C = 0.646$ Myr d$^{-1}$ and $k_I = 452$ d Myr$^{-1}$ listed there have been retained unmodified. This expression provides the age, $t$ (in Myr), explicitly in terms of the independent variables, $P$ and $\tau$.

All of the resulting individual ages lie between 2 and 3 Gyr, as can be seen in the histogram in Extended Data Fig. 9, which is peaked between these values. The formal mean age is 2.49 Gyr (1$\sigma$ = 0.25 Gyr), and the median age is 2.43 Gyr. The gyrochronology age for NGC 6819 is therefore 2.49 Gyr with a standard error of 0.056 Gyr (2%, ignoring possible systematic errors in gyrochronology).

## 4. Is contaminating light from close neighbours a problem?

NGC 6819 contains about 2,500 stars[25,37], is located 2.4 kiloparsecs (~7,800 light years) from the Sun[24,37–39] and is only 13.5 degrees above the Galactic plane. The cluster field is therefore densely populated with both cluster and foreground field stars. Accordingly, we must consider whether light from nearby stars can have 'leaked' into the photometric apertures used for our cluster targets. It would be difficult, if not impossible, to eliminate such contamination for all cluster stars with rotation period measurements. Instead, we take a qualitative approach to building a strong case against our overall result being unduly influenced by contaminating neighbours.

To accomplish this, we took advantage of our extensive stellar catalogue for the NGC 6819 field based on the deep, high-spatial-resolution CFHT/MegaCam images[25]. For each of 43 cluster stars for which we initially detected periodic photometric variability, we listed the astrometric and photometric properties for all neighbouring stars within a 20 arcsec radius. We also used the Kepler Guest Observer tool 'kepfield' to extract the pixel mask images (PMI) for each target star and for each quarter of observations. The 'kepfield' tool provides the coordinates, Kepler magnitudes and Kepler IDs for all stars in the Kepler Input Catalog (KIC) within the PMI. We searched for periodic variability in the light curves for all such neighbours observed by Kepler. Collectively, this information allowed us to study the angular separation, relative brightness and variability of neighbours within a ~5 pixel radius from each of the variable cluster stars. For the brightest neighbours, the photometric colour indices, effective temperatures and surface gravities derived from our star catalogue or given in the KIC, or both, provided additional guidance regarding their spectral type and evolutionary state.

From the 43 stars of interest, we rejected 13. Some of these stars were removed because the detected photometric variability was equal in period and in phase to variability detected in a close neighbour. Others were removed because of correlations between the presence and amplitude of periodic variability and the shape and size of the photometric aperture. Any such correlation suggests that the source of the variable signal originates outside the aperture and that a change in the aperture's size, shape and location results in more or less of the contaminating light being included. For some of the discarded stars the shape and size of the photometric aperture used in a given quarter was clearly affected by the signal from neighbouring stars. For others their location was significantly offset from the pixel with maximum counts inside the aperture, or, in the most severe cases, was entirely outside the aperture in one or more quarters. Extended Data Fig. 7 shows the PMIs for three quarters (Q) for one accepted star and one quarter each for two of the rejected stars. Extended Data Fig. 7a displays the PMI for Q15, Q16 and Q17 for accepted star KIC 4938993 (green circle within optimal aperture). This ~1.2 solar-mass star (V = 15.7 mag, $(B-V)_0$ = 0.5 mag) has 13 neighbours within 20 arcsec. The brightest neighbour is 1 mag brighter than KIC 4938993 and is located 19.6 arcsec away and outside the PMI. All other neighbours are 4 arcsec or more distant and 2.8 mag or more fainter. None of these neighbours has a light curve available from the Kepler archive. KIC 4938993 dictates the shape, location and size of the optimal aperture and falls on the optimal-aperture pixel with maximum counts in all quarters. It is thus highly unlikely that the 11.89 d, ~30 mmag amplitude signal observed for this star originates in a neighbouring star. In Extended Data Fig. 7b the PMI for quarter 8 for rejected star KIC 5023712 is displayed. The KIC star near the left edge of the PMI is 3.2 mag brighter than KIC 5023712, and was not observed by Kepler. The star at (column, row) = (592.1, 853.8) is 0.43 mag fainter and it is closer to the optimal aperture of KIC 5023712 than the 2 pixel radius of the circular aperture that captures 95% of its signal. This star varies with the same period and three times the amplitude of KIC 5023712, suggesting that the periodic signal observed for KIC 5023712 originates in this fainter neighbour. Extended Data Fig. 7c shows the PMI for quarter 5 for rejected star KIC 5287900 and how a nearby star brighter by 4.7 mag causes the optimal aperture to shift off the target entirely.

Following this analysis, we were left with 30 stars for which we believe the periodic variability in their light curves reflects rotational modulation. Our confidence in the 30 rotation periods is bolstered by the comparison between their expected projected rotation velocities and their spectroscopically measured $v\sin(i)$ values (Extended Data Fig. 8). Furthermore, the narrow rotational sequence traced by the 30 stars in the NGC 6819 CPD (Fig. 2) is by itself strong evidence against significant contamination. Neighbouring stars are likely to have masses (colours) or ages, or both, different from those of our target stars, and significant contamination would broaden the observed narrow sequence.

## 5. Are tidal interactions with close companions a concern for gyrochronology?

The gyrochronology age of a cool star is derived under the assumption that its rotational evolution has not been influenced by external forces. Although this is the case for the vast majority of cool Galactic field stars, a small fraction of stars will have a stellar or planetary companion close enough that tidal forces can drive an exchange of spin and orbital angular momentum between the two objects[40,41]. We provide here an estimate of the fraction of cool Galactic field stars for which tidal interactions can potentially affect their rotation.

The tidal torque scales as the inverse of the binary semi-major axis to the sixth power, restricting significant tidal evolution for unevolved cool stars to the very closest systems. For stellar binaries with solar-type components, tidal synchronization will have a significant effect on the stellar rotation only in systems with orbital periods less than ~20 d (semi-major axis less than ~0.18 AU; ref. 32). Using the distribution of orbital periods for field binaries with solar-type components[42], we find that binaries with periods of less than 20 d correspond to ~5% of that population. The same study[42] estimates that less than half (46%) of all solar-type stars are in binaries, implying that only about ~2.5% of cool Galactic field stars have a stellar companion close enough for effective tidal interactions.

Short-period planetary companions can also tidally interact with their hosts, causing an increase in the stellar rotation period. The relevant class of planets here is the hot Jupiters, that is, planets with at least 10% of Jupiter's mass and orbits of less than 10 d (ref. 43). Such planets occur around solar-type field stars with a frequency of ~1% (ref. 43). Although the efficiency of tidal evolution in such star–planet systems is still uncertain on theoretical and observational grounds, it is probably only the hottest and most massive of the hot Jupiters that will have a significant effect on the rotational evolution of their host stars.

Assuming for simplicity that hot Jupiters do not occur in binaries with orbital periods shorter than 20 d, we estimate that at most 3% of cool stars have stellar or planetary companions that can affect their rotation through tidal interactions. Thus, tidal effects on rotation are not a concern for ~97% of cool Galactic field stars.

# EXTENDED DATA

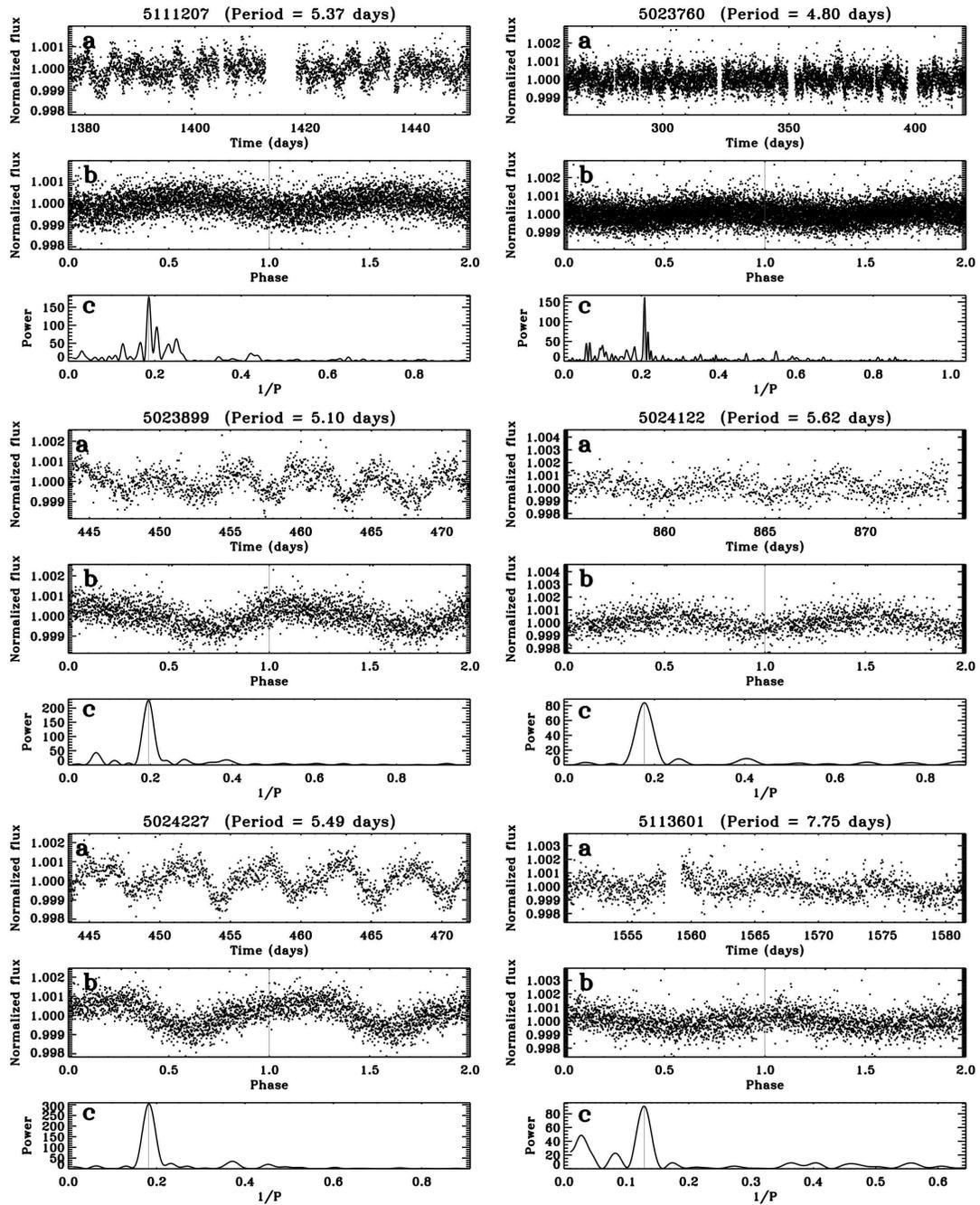

**Extended Data Figure 1. | The light curves, phase folded light curves, and periodograms for the stars KIC 5111207, 5023899, 5024227, 5023760, 5024122, 5113601.** For each of the 30 stars we show a segment of the full *Kepler* light curve used for determining its rotation period (**a**), the corresponding phase folded light curve (**b**), and the periodogram (power as a function of rotation frequency, **c**). The KIC identification number [36] and the measured rotation period for each star is shown above the light curve segment in the top panel.

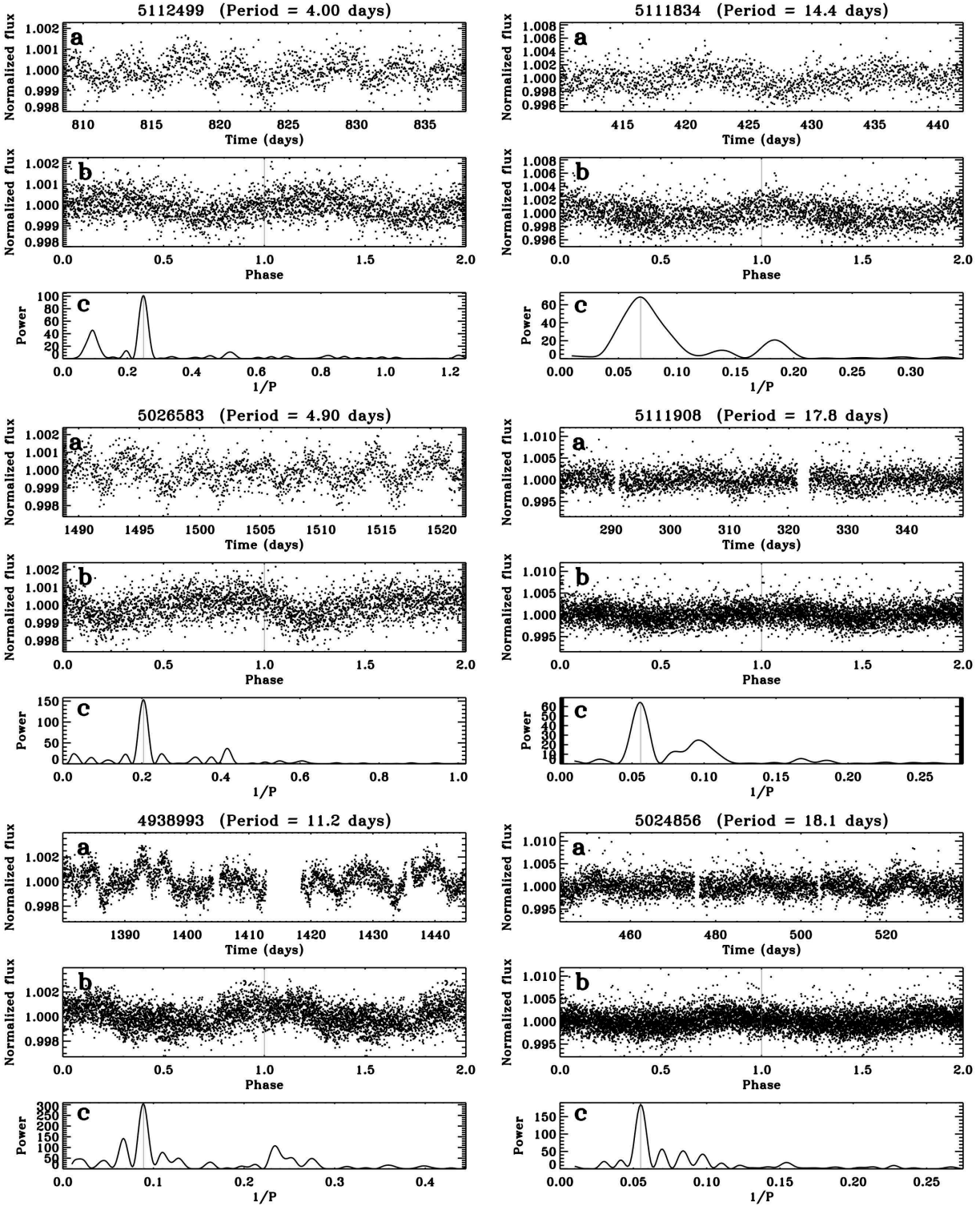

**Extended Data Figure 2. | The light curves, phase folded light curves, and periodograms for the stars KIC 5112499, 5026583, 4938993, 5111834, 5111908, 5024856.** See Extended Data Fig. 1 for details.

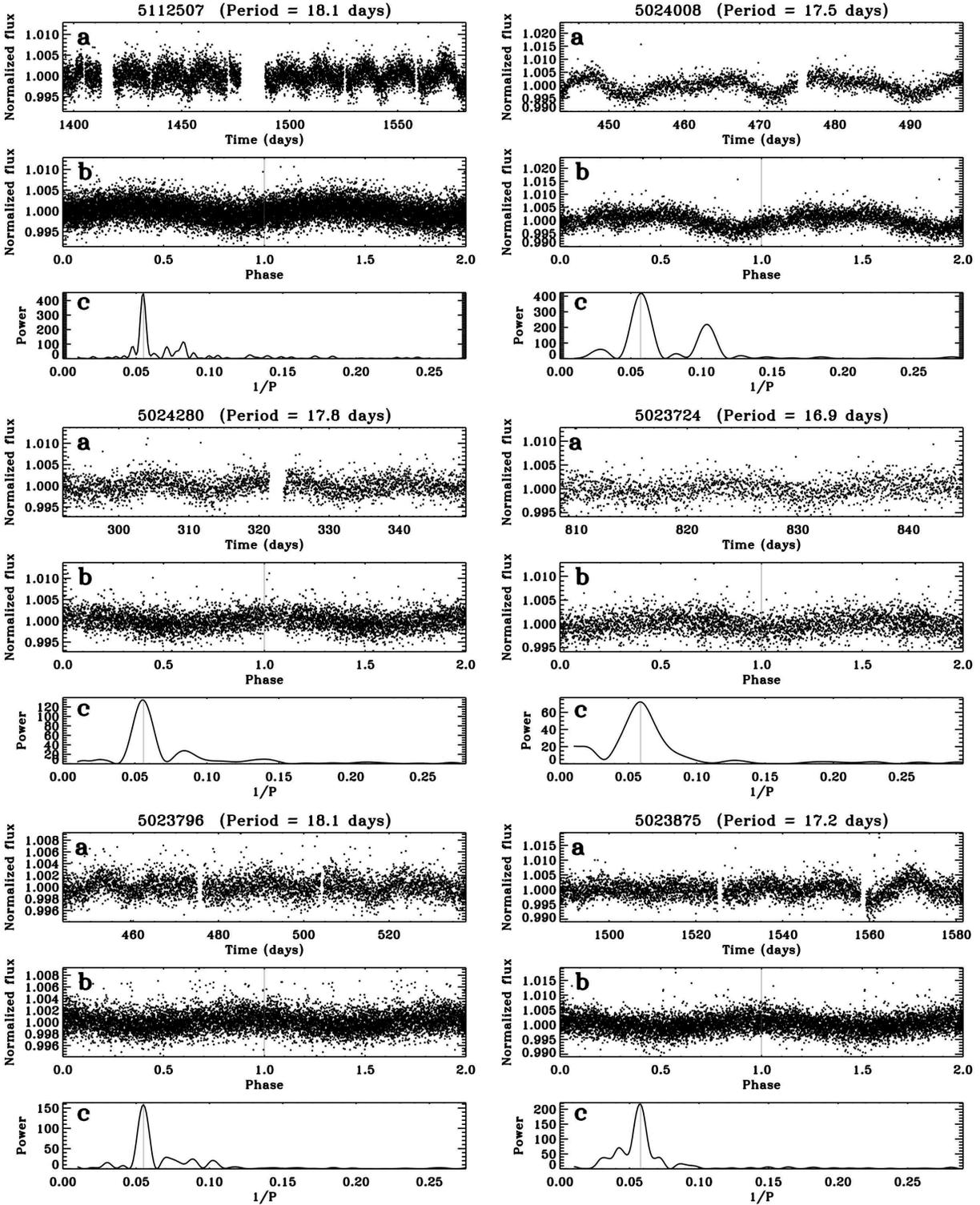

**Extended Data Figure 3. | The light curves, phase folded light curves, and periodograms for the stars KIC 5112507, 5024280, 5023796, 5024008, 5023724, 5023875.** See Extended Data Fig. 1 for details.

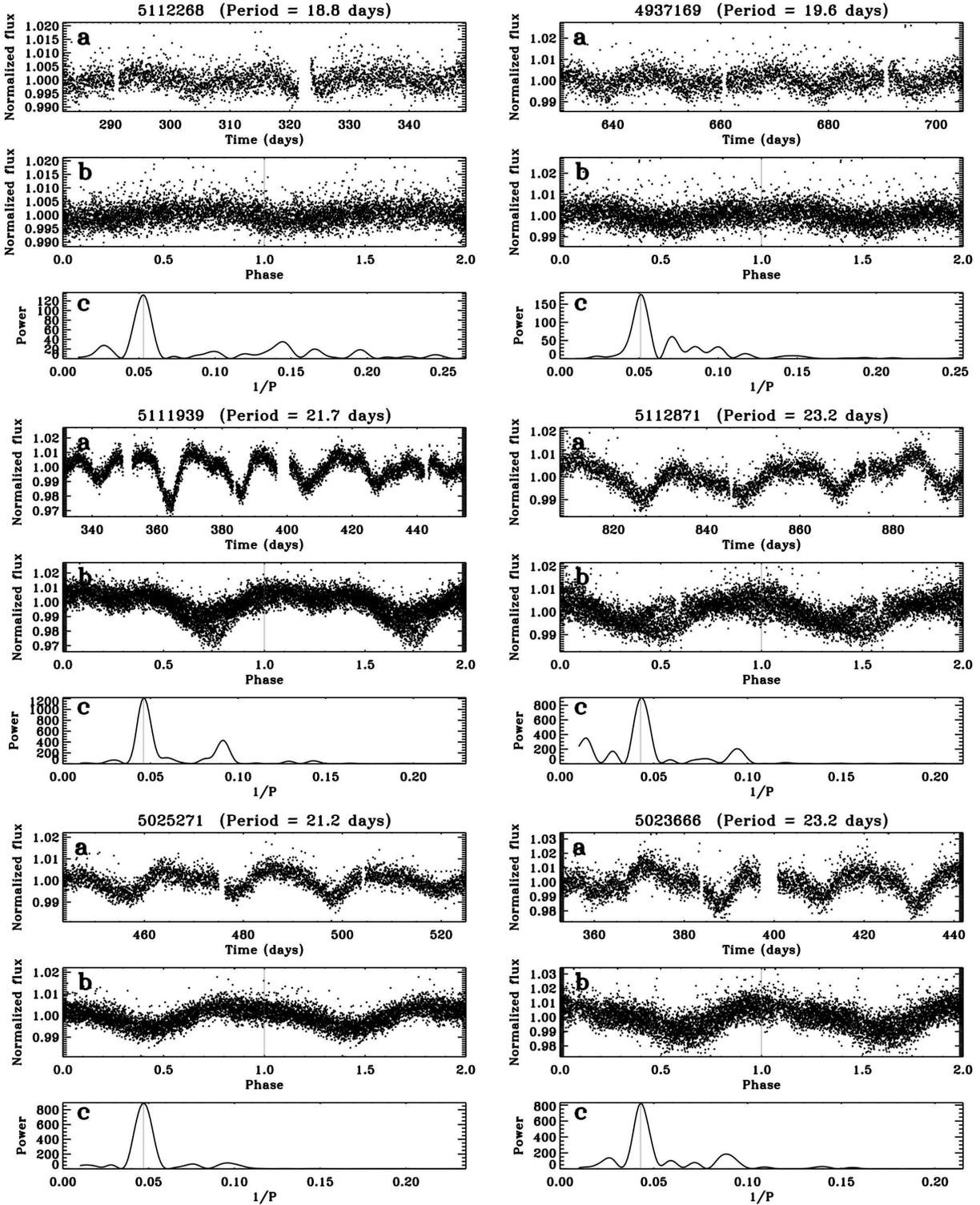

**Extended Data Figure 4. | The light curves, phase folded light curves, and periodograms for the stars KIC 5112268, 5111939, 5025271, 4937169, 5112871, 5023666.** See Extended Data Fig. 1 for details.

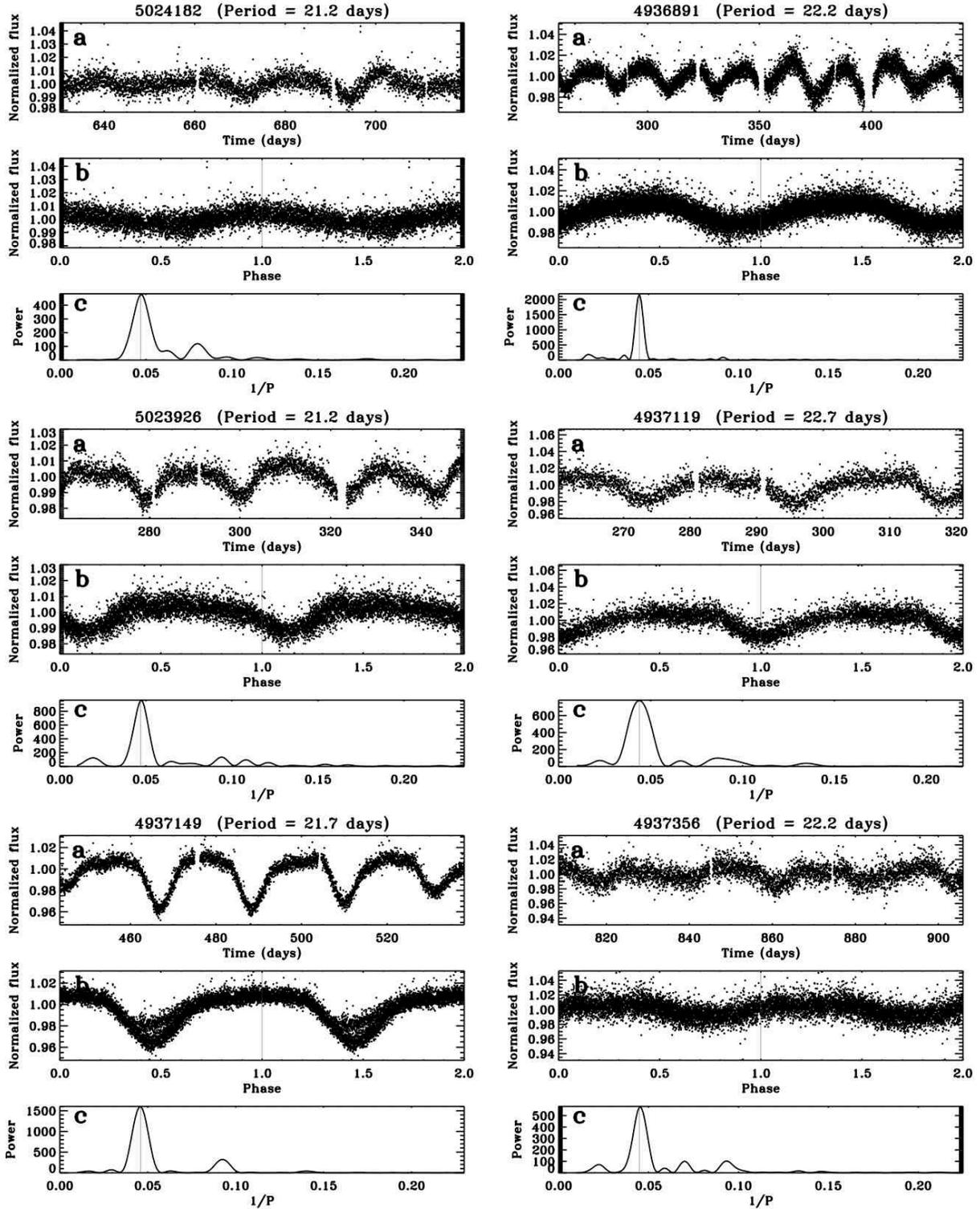

**Extended Data Figure 5. | The light curves, phase folded light curves, and periodograms for the stars KIC 5024182, 5023926, 4937149, 4936891, 4937119, 4937356.** See Extended Data Fig. 1 for details.

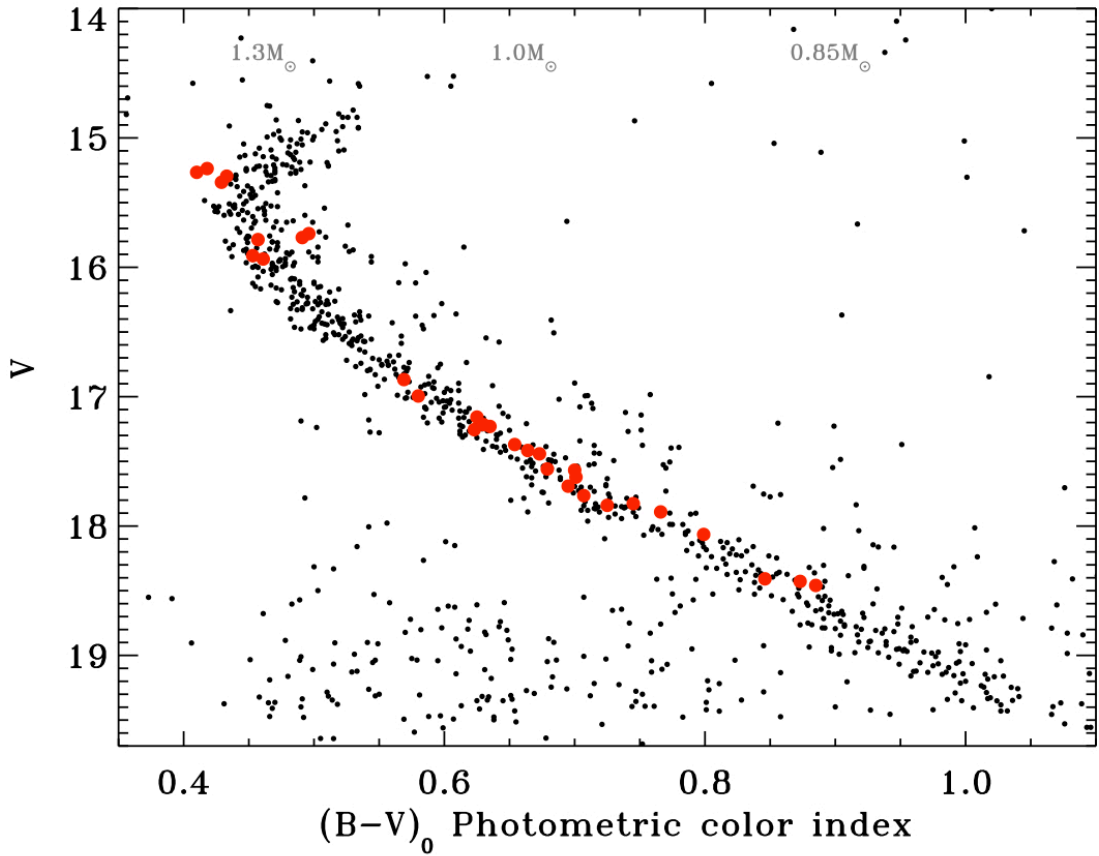

**Extended Data Figure 6. | The NGC 6819 color-magnitude diagram.** The color-magnitude diagram for stars identified as a common proper motion members of NGC 6819 [25] and located within 5 arcminutes from the cluster center. The diagonal band tracing a tight relationship between the de-reddened photometric color index, $(B-V)_0$, and brightness, V, represents the population of cluster members. The locations of the 30 stars with measured rotation periods are marked with larger red circles. They all fall along this band and are thus photometric members of NGC 6819. Stellar masses in solar units are given along the top horizontal axis at the corresponding colors. The light from distant binary companions causes the two rotators near $(B-V)_0$ of 0.5 to fall above the cluster sequence.

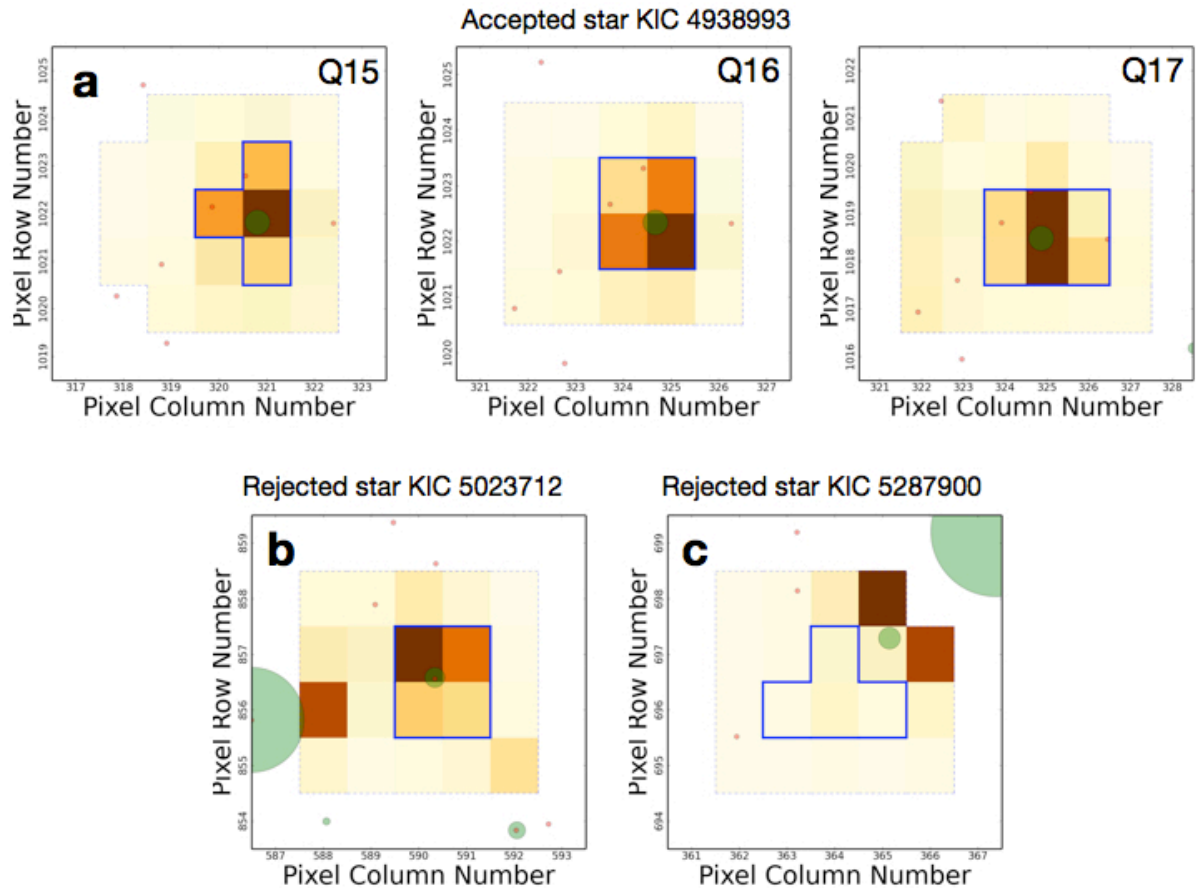

**Extended Data Figure 7. | Pixel mask images for NGC 6819 members.** Examples of pixel mask images (PMI) for the accepted star KIC 4938993 (**a**) and the rejected stars KIC 5023712 and KIC 5287900 (**b**,**c**). Transparent green circles mark the positions of KIC sources. Red dots correspond to the positions of fainter sources from deeper surveys within the *Kepler* field. The solid blue line traces the optimal aperture (OA; optimizing the signal to noise ratio for the target) defined for each target star in each quarter. The shape, size, and location of the OA was typically different for the different quarters of observations.

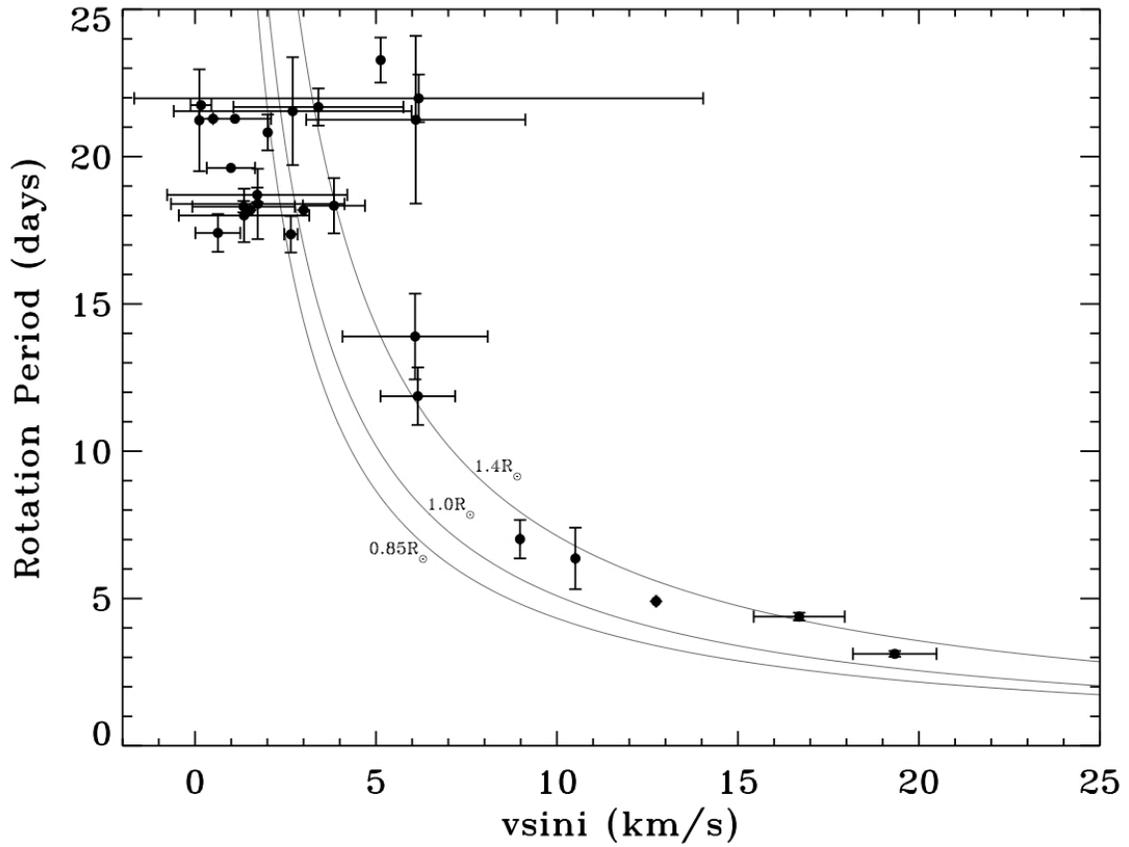

**Extended Data Figure 8. | A comparison of rotation periods and projected rotation velocities for stars in NGC 6819.** Projected rotation velocities (*vsini*) plotted against the measured rotation periods for stars in NGC 6819. For comparison, three solid black curves show the expected relations between rotation period and *vsini* for stars with radii of 0.85, 1.0, and 1.4 times solar, observed at an inclination of 90 degrees. All stars plotted have single-lined spectra. The average velocity resolution in the Hectochelle spectra is 7.38 km s$^{-1}$. The agreement between the expected and observed *vsini* values for the measured rotation periods provide additional validation of our rotation period measurements.

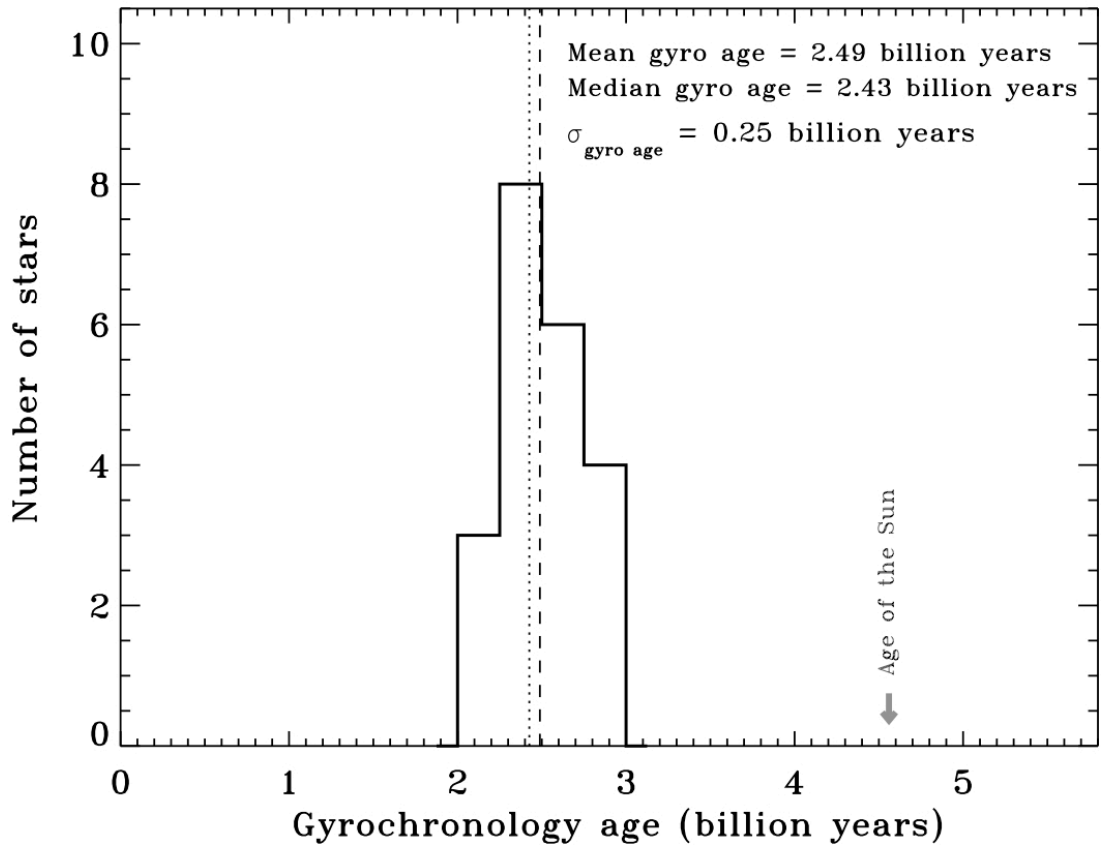

**Extended Data Figure 9. | The gyro-age distribution for 21 cool dwarf members of NGC 6819.** The gyrochronology ages for the 21 stars in the NGC 6819 CPD with $(B-V)_0$ colors in the range from 0.55 to 0.9 (masses between ~1.1 and 0.85 solar). The mean and median of the distribution are 2.49 and 2.43 billion years, respectively. The standard deviation for the 21 gyro-ages is 0.25 billion years, or 10% of the mean gyro-age for the cluster.

**Extended Data Table 1 | Basic parameters and rotation period measurements for 30 members of NGC6819**

| Kepler ID | Right Ascension (hr min sec) | Declination (° ′ ″) | V (magn) | $(B-V)_0$ (magn) | nP | $P_{Mean}$ (Days) | $\sigma_P$ (Days) | $P_{RV}$ (%) | $P_\mu$ (%) |
|---|---|---|---|---|---|---|---|---|---|
| 5111207 | 19 40 09.01 | 40 12 16.7 | 15.27 | 0.41 | 7 | 5.42 | 1.05 | 92 | 61 |
| 5023899 | 19 40 55.90 | 40 10 04.7 | 15.24 | 0.42 | 6 | 4.81 | 0.44 | 92 | 99 |
| 5023760 | 19 40 49.33 | 40 09 07.0 | 15.30 | 0.43 | 6 | 4.78 | 0.10 | 75 | 99 |
| 5024227 | 19 41 09.49 | 40 11 41.6 | 15.34 | 0.43 | 3 | 5.06 | 0.39 | 88 | 99 |
| 5024122 | 19 41 05.50 | 40 08 28.4 | 15.91 | 0.45 | 2 | 6.36 | 1.04 | 93 | 96 |
| 5112499 | 19 41 13.91 | 40 15 30.3 | 15.94 | 0.46 | 3 | 4.44 | 0.92 | 93 | 99 |
| 5113601 | 19 42 00.18 | 40 12 19.2 | 15.79 | 0.46 | 5 | 7.01 | 0.65 | 92 | 99 |
| 5026583 | 19 42 55.25 | 40 09 48.0 | 15.77 | 0.49 | 1 | 4.90 | - | 91 | 36 |
| 4938993 | 19 42 55.85 | 40 00 19.8 | 15.74 | 0.50 | 3 | 11.87 | 0.98 | 81 | 62 |
| 5111834 | 19 40 43.86 | 40 15 18.4 | 16.87 | 0.57 | 7 | 13.89 | 1.46 | 92 | 97 |
| 5111908 | 19 40 43.86 | 40 12 53.2 | 17.00 | 0.58 | 2 | 17.41 | 0.64 | 93 | 97 |
| 5024856 | 19 41 30.61 | 40 08 42.8 | 17.26 | 0.62 | 1 | 18.19 | - | 76 | 95 |
| 5024280 | 19 41 11.63 | 40 08 34.1 | 17.22 | 0.63 | 3 | 17.36 | 0.62 | 93 | 96 |
| 5112507 | 19 41 14.09 | 40 17 34.7 | 17.16 | 0.63 | 1 | 18.19 | - | 91 | 97 |
| 5023796 | 19 40 51.10 | 40 09 45.1 | 17.23 | 0.64 | 3 | 18.30 | 0.19 | 67 | 96 |
| 5024008 | 19 41 00.51 | 40 10 29.4 | 17.37 | 0.65 | 2 | 18.40 | 1.19 | 93 | 96 |
| 5023724 | 19 40 47.50 | 40 07 40.3 | 17.42 | 0.66 | 3 | 18.00 | 0.91 | 91 | 92 |
| 5023875 | 19 40 54.82 | 40 09 20.5 | 17.44 | 0.67 | 3 | 18.33 | 0.94 | 77 | 95 |
| 5112268 | 19 41 04.25 | 40 14 23.2 | 17.56 | 0.68 | 2 | 18.70 | 0.25 | 91 | 94 |
| 4937169 | 19 41 22.61 | 40 05 44.6 | 17.62 | 0.70 | 1 | 19.62 | - | 93 | 95 |
| 5025271 | 19 41 47.15 | 40 09 30.3 | 17.57 | 0.70 | 2 | 21.29 | 0.10 | 89 | 93 |
| 5111939 | 19 40 49.63 | 40 12 05.2 | 17.70 | 0.70 | 1 | 21.75 | - | 92 | 90 |
| 5112871 | 19 41 26.64 | 40 16 01.4 | 17.77 | 0.71 | 2 | 21.25 | 2.85 | 78 | 86 |
| 5023666 | 19 40 44.23 | 40 10 11.6 | 17.84 | 0.73 | 3 | 21.54 | 1.83 | 92 | 88 |
| 5024182 | 19 41 07.57 | 40 08 50.3 | 17.83 | 0.75 | 1 | 21.29 | - | 90 | 92 |
| 5023926 | 19 40 56.89 | 40 06 10.1 | 17.89 | 0.77 | 4 | 20.82 | 0.61 | 93 | 91 |
| 4937149 | 19 41 21.54 | 40 05 21.1 | 18.07 | 0.80 | 6 | 21.68 | 0.63 | 92 | 88 |
| 4936891 | 19 41 07.75 | 40 05 55.0 | 18.41 | 0.85 | 7 | 21.98 | 0.81 | 92 | 82 |
| 4937119 | 19 41 19.90 | 40 05 39.7 | 18.43 | 0.87 | 2 | 23.28 | 0.77 | 89 | 59 |
| 4937356 | 19 41 33.09 | 40 05 12.4 | 18.46 | 0.89 | 3 | 21.23 | 1.73 | 93 | 78 |

The *Kepler* ID is from the Kepler Input Catalog (KIC; [36]). Right Ascension and Declination are equinox J2000 [25]. The stellar brightness, V, and color index, $(B-V)_0$, are based on CCD photometry using the CFH12K mosaic CCD on the 3.6m Canada-France Hawaii Telescope (CFHT) telescope [24]. The B-V colors were dereddened using a value for the color excess $E_{(B-V)}$ of 0.15 [37,38,39]. For stars with more than one rotation period measurement (nP > 1), the $P_{Mean}$ and $\sigma_P$ columns list the mean period and the standard deviation of the multiple measurements.